Primitive chain network simulations of the creep of entangled polymers

Yuichi Masubuchi[1], Giovanni Ianniruberto[2], and Giuseppe Marrucci[2]

1) Department of Materials Physics, Nagoya University, Nagoya 4648603, Japan

2) Dipartimento di Ingegneria Chimica, dei Materiali e della Produzione Industriale, Università degli Studi di Napoli "Federico II", Piazzale Tecchio 80-80125 Napoli, Italy

ABSTRACT

Although the behavior of entangled polymers in startup shear flows with constant shear rates has been thoroughly investigated, the response under creep has not been frequently considered. In this study, primitive chain network simulations, based on a multi-chain sliplink model, are modified so as to describe creep experiments. Creep simulations are compared to a literature dataset of an entangled polybutadiene solution, and qualitative agreement is found in the nonlinear range, i.e., under large stresses. Simulations allow one to extract details of the transient molecular motion, and results suggest that the deformation-induced disentanglement is relatively mild in the stress-controlled mode as compared to the rate-controlled one, because coherent molecular tumbling at the start of flow is disrupted.

KEYWORDS

Coarse-grained molecular simulations, entangled polymers, shear flows, creep compliance

INTRODUCTION

Creep measurements of polymeric liquids are extensively present in the literature. For instance, Aiken et al. [1] reported the tensile creep behavior of plasticized polyvinyl chlorides. Bueche [2] investigated the rheology of polymethylmethacrylate and methyl-decyl copolymer melts by tensile creep. Similar experiments have been conducted for several other polymers, including creep in shear [3,4] instead of the usual tensile test. Attempts have also been made to model the observed behavior. Tshoegl [5] applied Rouse and Zimm models to creep. Watanabe and Inoue [6] published a review on predictions of the Rouse model for molecular conformations under creep and recovery. Attempts to model entangled polymers have also been reported, although the approach was essentially phenomenological. [2,7–11]

Although creep response is no longer at the forefront of polymer research, nonlinear creep, i.e., under large stresses, remains worth being investigated. Wang and coworkers [12–14] studied the nonlinear creep of entangled polybutadiene solutions, and they reported a yield-like behavior, i.e., the shear rate steeply increases in a narrow range of applied shear stress. They argue that the phenomenon reflects chain disentanglement, and they refer to this behavior as the entanglement-disentanglement transition (EDT).

Change in entanglement density has been experimentally studied through determination of the plateau modulus under deformation, i.e., via superposition measurements [15,16]. Molecular simulations of entangled polymers have also been used to determine the entanglement density in the network [17]. However, to our knowledge, such analyses have never been attempted for creep tests. Stress-controlled molecular simulations have been rarely performed, as shear rate-controlled ones are simpler.

In this study, the primitive chain network simulation [18,19] is extended to the creep mode. After confirming qualitative agreement with the experiments by Ge et al. [14], molecular details, including the entanglement density, are analyzed. Results reveal that deformation-induced disentanglement is observed both for rate-controlled and stress-controlled simulations when the shear rate or external stress is significant, i.e., in the nonlinear range. However, disentanglement is somewhat suppressed in the stress-controlled case, due to the disrupted coherence in molecular tumbling. Details are shown below.

MODEL AND SIMULATIONS

Since the model used in this study has been previously reported, only a brief explanation is given below. The entangled polymeric liquid is replaced by a network consisting of nodes, strands, and dangling ends. Each polymer chain corresponds to a path connecting two dangling ends via several nodes and strands. At each node, two polymer chains entangle with one another (binary assumption), and a sliplink is located to bundle the pair of segments.

The position of the sliplinks obeys a Langevin-type equation of motion, where a balance is considered among the drag force, the tensions acting in the chain strands, the osmotic force (controlling density fluctuations), and a thermal random force:

$$2\zeta\left(\frac{d\mathbf{R}}{dt} - \boldsymbol{\kappa}\cdot\mathbf{R}\right) = \frac{3kTn_0}{a^2}\sum_i^4 \frac{\mathbf{r}_i}{n_i} - n_0\nabla\mu + \mathbf{F}_B \qquad (1)$$

Here, $\zeta$ is the friction coefficient of each strand, $\mathbf{R}$ is the sliplink position, and $\boldsymbol{\kappa}$ is the velocity gradient tensor of the flow, assumed homogeneous in the entire system. Friction changes under fast flows [20,21] are not considered in this study. Symbols $a$ and $n_0$ refer to the average strand length and average Kuhn segment number in a strand under equilibrium, respectively, while $\mathbf{r}_i$ and $n_i$ are the strand vector and the Kuhn segment number in the strands connected to the sliplink. Finite chain extensibility [22] is neglected for simplicity. According to the binary assumption of entanglement mentioned above, there are four connected strands for each sliplink. Symbol $\mu$ refers to the chemical potential defined from the free energy written below[19,23]:

$$A = \begin{cases} \varepsilon\left(\frac{\phi(\mathbf{R})}{\langle\phi\rangle} - 1\right)^2 & \text{for } \phi(\mathbf{R}) > \langle\phi\rangle \\ 0 & \text{for } \phi(\mathbf{R}) \leq \langle\phi\rangle \end{cases} \qquad (2)$$

Here, $\phi(\mathbf{R})$ is the local sliplink density, and $\langle\phi\rangle$ is its average for the entire system. Going back to eq 1, $\mathbf{F}_B$ is the thermal random force obeying the fluctuation-dissipation theorem, i.e., $\langle\mathbf{F}_B\rangle = \mathbf{0}$ and $\langle\mathbf{F}_B(t)\mathbf{F}_B(t')\rangle = 4\zeta\delta(t-t')\mathbf{I}$. Possible changes of the Brownian force intensity in fast flows [24] are not considered.

Chain sliding through the sliplinks (reptation) is described by a 1D Langevin equation giving the rate of change for the number of Kuhn segments in a strand:

$$\frac{\zeta}{\varphi}\frac{dn_{i+1}}{dt} = \frac{3kTn_0}{a^2}\left(\frac{r_{i+1}}{n_{i+1}} - \frac{r_i}{n_i}\right) - n_0\nabla\mu + f \qquad (3)$$

More specifically, this equation gives the change rate of $n_{i+1}$, due to the transport of Kuhn segments from the connected strand $i$. The equation is the 1D force balance along the chain, with forces similar to those considered for the 3D sliplink position. Symbol $\varphi$ refers to the local linear density of Kuhn segments:

$$\varphi = \frac{1}{2}\left(\frac{n_{i+1}}{r_{i+1}} + \frac{n_i}{r_i}\right) \quad (4)$$

In eqs 3 and 4, $r_i$ is the strand length, and $f$ is the 1-D random force, obeying $\langle f \rangle = 0$ and $\langle f(t)f(t') \rangle = 2\zeta\delta(t-t')$.

In addition to the dynamics of **R** and $n$, the creation and destruction of sliplinks at the chain ends occur when the number $n_e$ of Kuhn segments at the dangling ends exceeds the range $1/2 \leq n_e/n_0 \leq 3/2$ [25,26]. To create a new sliplink, the protruding, dangling end hooks a randomly chosen chain strand within a given distance. Vice versa, when a dangling end slides off, the connected sliplink is eliminated. These changes in network topology are locally equilibrated (according to the dynamics mentioned above) within the relaxation time of the strand given by $\tau = \zeta a^2/6kT$.

The simulations were performed with non-dimensional quantities, by using the average strand length $a$, the thermal energy $kT$, and the strand relaxation time $\tau$ as units of length, energy, and time, respectively. The number of Kuhn segments in each strand was normalized with its average value $n_0$. When comparing simulation results with experiments, however, units of modulus $G_0 = kT/a^3$ and molecular weight $M_0$ were employed [27,28], in place of $a$ and $n_0$. The polymer chains were dispersed in a simulation box with a Kuhn segment density of $10n_0/a^3$ (i.e., 10 in nondimensional units). The box dimensions were $(12a)^3$ and $(16a)^3$ for simulations of linear and nonlinear viscoelasticity, respectively. Usual periodic boundary conditions were adopted.

To perform creep simulations, we here adopt the following procedure. The shear stress is calculated from the strand vector $\mathbf{r}_i$ between two adjacent sliplinks and the corresponding Kuhn segments number $n_i$ by using the well-known Kramer's expression:

$$\sigma_{xy} = \frac{3kT}{V} \sum_i \left(\frac{r_{xi} r_{yi}}{n_i}\right) \quad (5)$$

Here, $V$ is the volume of the simulation box, and the subscripts $x$ and $y$ refer to the shear and shear gradient components of $\mathbf{r}_i$. During the simulations, $\sigma_{xy}$ is monitored, and the shear rate $\dot{\gamma}$ of the background flow is controlled to realize a $\sigma_{xy}$ value close to the set value $\sigma_r$ as follows:

$$\dot{\gamma}(t + \tau) = \dot{\gamma}(t) \left(\frac{\sigma_r}{\sigma_{xy}(t)}\right)^\alpha \quad (6)$$

Here, $\alpha$ is a parameter that determines the stiffness of this feedback control, and it was empirically chosen to be $\alpha = 16$. A similar scheme was employed in the previous study on wall-slip[29].

Simulation results are then compared with creep experiments on an entangled polybutadiene (PB) solution performed by Ge et al. [14]. The PB molecular mass $M$ is 1.05x10³ kg/mol, and the "solvent" is a butadiene oligomer with a molar mass of 10.5 kg/mol. The polymer concentration is 5%, and the temperature $T$ is 27°C. From the plateau modulus $G_N = 1.3$ kPa of the PB solution, the entanglement molecular weight $M_e = \rho RT / G_N$ is calculated as 86 kg/mol ($\rho$ being the polymer density, $\rho = 45$ kg/m³). According to the empirical relation established in previous studies [30], $M_0 = 2M_e/3$, we find $M_0 = 57$ kg/mol. Hence, the average number of entangled strand segments at equilibrium is $Z = M/M_0 = 18$. Figure 1 shows a typical snapshot of the simulation box in a steady shear flow with a shear rate of $0.037\tau$, demonstrating that the simulation box is sufficiently larger than the molecule.

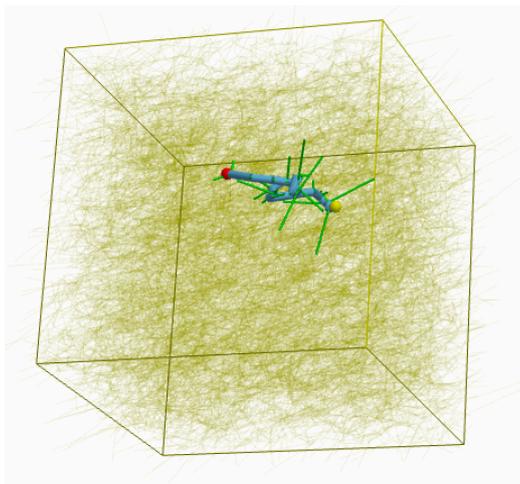

**Figure 1** A typical snapshot of the $(16a)^3$ simulation box in a steady shear flow at a shear rate of $0.037/\tau$. One of the 2,276 molecules is highlighted, thin yellow lines showing the others. Red and yellow spheres indicate chain ends. Green thin lines are strands entangled with the highlighted molecule.

RESULTS AND DISCUSSION

Figure 2 reports the comparison between linear viscoelastic data and simulation results. The linear relaxation modulus was calculated in the simulations from the stress-autocorrelation by using the Green-Kubo formula, and then converted to $G'$ and $G''$ through the reptate software [31]. As reported earlier [28], the simulated linear viscoelastic response is in good agreement with experiments, except for the high-frequency behavior, which corresponds to higher Rouse modes and glassy modes, not accounted for in our coarse-grained simulations. Mapping of simulation results onto experimental data gives $G_0 = 2.9$ Pa and $\tau = 0.1$ s. We employ these parameter values for nonlinear viscoelastic simulations. It is fair to note that, for unknown reasons, this $G_0$ value is somewhat inconsistent with the value $G_0 = 2.0$ Pa obtained from the relation $G_0 = \rho RT/M_0$.

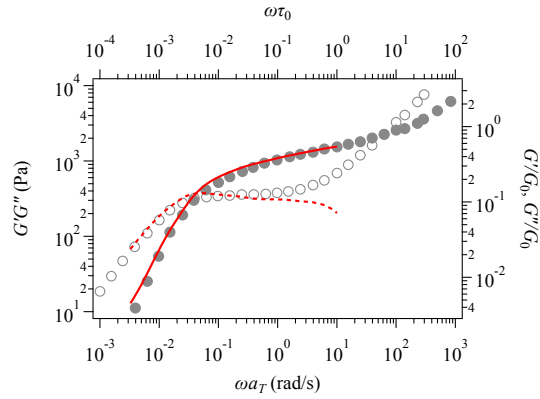

**Figure 2** Linear viscoelastic response of the PB solution. Symbols are data from [14], lines are our simulation results.

Figure 3 shows the viscosity growth curves in shear-rate-controlled shear flows. As reported earlier, the simulation reasonably reproduces the data, including shear thinning and stress overshoots.

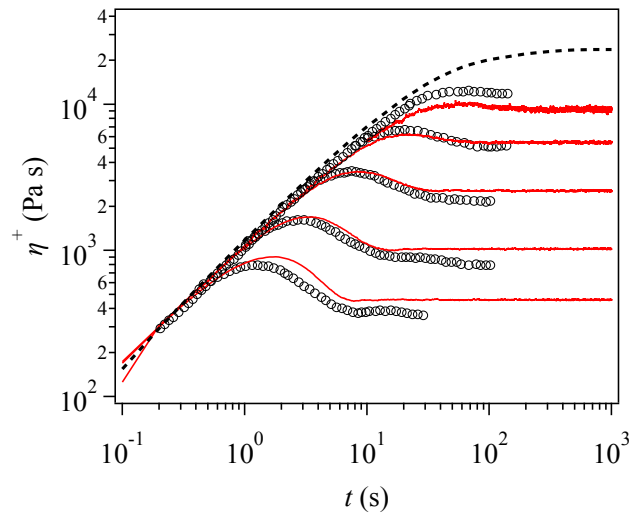

**Figure 3** Startup viscosity growth in shear-rate-controlled shear flows, at shear rates equal to 3, 1, 0.3, 0.1, and 0.03 s$^{-1}$, from bottom to top. Symbols are data from [14]. Red curves are simulation results. The broken (black) curve is the linear viscoelastic envelope.

Figure 4(a) refers to the creep compliance. The simulation results reasonably agree with the experimental data when the applied stress is relatively small (see green and black symbols and curves). In contrast, the data under large stress (red and blue symbols) are not well reproduced by the simulations. Indeed, although simulations nicely capture the data at short times, including the undershoot, they reach the steady state earlier than the experiments and, consequently, underestimates the steady-state shear rate. The discrepancy becomes more apparent when plotting the shear rate normalized by the applied stress, as shown in Fig. 4(b).

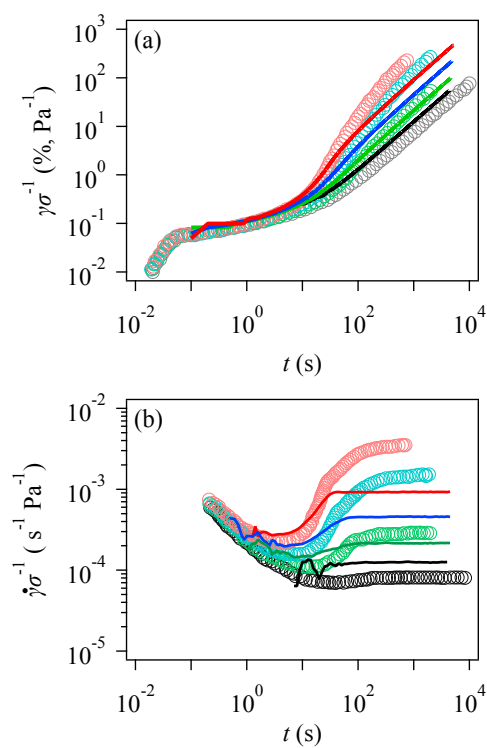

**Figure 4** Time development of shear strain (a) and shear rate (b) normalized by the applied stress at 1000 (red), 800 (blue), 600 (green), and 370 Pa (black), respectively. Symbols are experiment from [14]. Curves are simulation results.

Figure 5 shows the flow curve, i.e., the plot shear stress vs. shear rate at steady state. This plot explains why the simulation somewhat fails to capture the creep behavior in Fig. 4, despite reasonably reproducing the startup shear data in Fig. 3. Specifically, in Fig. 5, at each given shear rate, shear stress predictions are quite consistent with data, i.e., the vertical distance is not significant. In contrast, at a given shear stress, shear rate predictions are quite far from data, i.e., the horizontal distance between simulation results and data is significant. This is due to the fact that in the high shear rate region there is a power-law dependence of the shear stress on the shear rate, with an exponent of ca. 0.24. In this respect, tests of simulations (or models) in stress-controlled experiments are more severe than rate-controlled ones. Since the simulated flow curve can overlap with data by an arbitrary shift on the logarithmic plot, agreement for the creep behavior could be improved if the parameters are further optimized, albeit at the expense of the agreement for the linear viscoelasticity in Fig. 2. Specifically, the linear response in the low-shear rate regime is underestimated in Fig. 5, despite the agreement in Fig. 2.

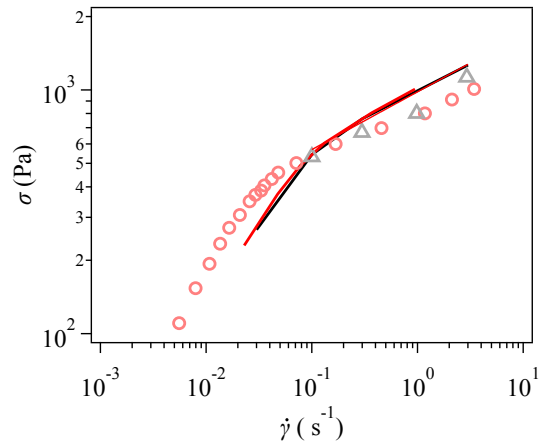

**Figure 5** Shear stress vs. shear rate for rate-controlled (red) and stress-controlled (black) steady shear flows. Symbols are experiments from [14], curves are simulation results.

Although the simulation captures the experiments only qualitatively, it proves instructive to discuss the molecular behavior by invoking the following stretch-orientation decoupling approximation [32,33].

$$\boldsymbol{\sigma} \approx 3\left(\frac{Z}{Z_0}\right)\lambda^2 \mathbf{S} \qquad (7)$$

Here, $Z$ is the number of entangled strands per molecule, $\lambda$ is the strand stretch, and $\mathbf{S}$ is the strand orientation tensor. Figure 6 shows the time development of these quantities for the stress-controlled and the shear-rate-controlled modes, attaining a similar steady-state shear rate and shear stress: $\sigma_{xy}/G_0 = 0.28$ and $\dot{\gamma}\tau = 3.7 \times 10^{-2}$. Figure 6(a) refers to $\sigma_{xy}(t)$, with and without the decoupling approximation. For both rate-controlled and stress-controlled cases, the approximation works reasonably well, although the steady-state stress is slightly overestimated when using the approximation. Figure 6(b) shows $\dot{\gamma}(t)$, which exhibits an undershoot for the stress-controlled case (red curve). Ge et al. [14] discussed the relation between this undershoot in $\dot{\gamma}(t)$ for the stress-controlled case and the overshoot in $\sigma_{xy}(t)$ for the rate-controlled one. Consistently with their discussion, the positions of the stress maximum and the shear rate minimum coincide.

As regards the separate role of stretch and orientation, rate-controlled simulations exhibit the established behavior [32,34,35]. Namely, the stress overshoot is mainly attributable to the segment orientation $S_{xy}$ (Fig. 6e) and delayed by the subsequent segment stretch $\lambda^2$ (Fig. 6d). A reduction of $Z$ (Fig. 6c) due to convective constraint release [36,37] also contributes to the decrease of $S_{xy}$ after the overshoot. In the stress-controlled simulation, $S_{xy}$ rises immediately to a specific value to sustain the required stress. It slightly decreases to compensate for the growth of $\lambda^2$ due to the reduction of $Z$. These changes, indicated by the red curves, are monotonic, i.e., without any overshoot or undershoot, despite the undershoot in $\dot{\gamma}$. Consequently, they are relatively gradual compared to the rate-controlled case, shown by the black curves.

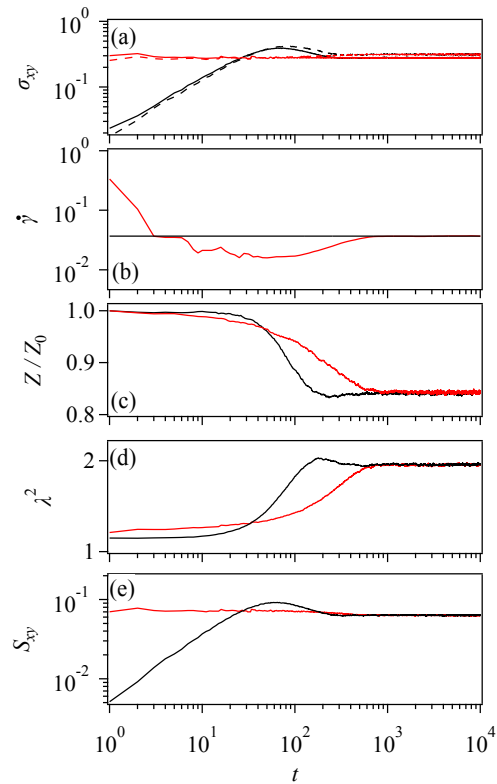

**Figure 6** Time development of $\sigma_{xy}$ (a), $\dot{\gamma}$ (b), $Z/Z_0$ (c), $\lambda^2$ (d), and $S_{xy}$(e), for the rate-controlled (black) and stress-controlled (red) simulations, with a similar steady-state with $\sigma_{xy} =$

0.28 and $\dot{\gamma} = 3.7 \times 10^{-2}$. Broken curves in panel (a) show results with the decoupling approximation.

As demonstrated by the decoupling analysis in Fig. 6, the transient behavior before the steady state differs between the rate-controlled and stress-controlled cases. This difference is related to the coherent molecular tumbling at the start of flow [38–42]. Figure 7(a) shows the time development of the square value of the shear gradient component $u_y^2$ of the orientation **u** of the chain end-to-end vector. For the rate-controlled cases (shown by black curves), $u_y^2 \sim 1/3$ at $t = \tau$ since the chains are randomly oriented ($u_x^2 + u_y^2 + u_z^2 = 1$). As the strain accumulates, the chains tilt toward the shear direction, and $u_y^2$ decreases to a specific steady value depending on $\dot{\gamma}$. Before reaching the steady state, $u_y^2$ exhibits an undershoot when $\dot{\gamma}$ is large. This $u_y^2$ response reflects the tumbling of molecules in a coherent manner, as discussed earlier [39]. For the stress-controlled cases (indicated by red curves), $u_y^2$ is smaller than $1/3$ at $t = \tau$ since $S_{xy} > 0$ to sustain the required stress, as shown in Fig. 6(e). Then $u_y^2$ decreases toward the steady value, which is irrespective of the control mode. However, in the stress-controlled case, the reduction of $u_y^2$ is monotonic, i.e., without any undershoot, implying that the coherence in molecular tumbling is suppressed.

Concerning the entanglement-disentanglement transition (EDT), Figure 7(b) shows the surviving fraction of sliplinks $v_s$. (Note that surviving means the sliplinks that remain from the start of the flow, whereas $Z$ shown in Fig. 6(c) is the current number of sliplinks, including those created or destroyed during the flow.) The fraction $v_s$ decays faster under higher $\dot{\gamma}$ and $\sigma_{xy}$. Thus, deformation-induced disentanglement is observed, irrespective of the deformation mode. However, for similar steady states, the decay of $v_s$ is faster in rate-controlled cases than in stress-controlled ones. The decay of $v_s$ is consistent with the behavior of $u_y^2$, as tumbling hardly occurs unless $v_s$ sufficiently decreases. These observations suggest that deformation-induced disentanglement is suppressed under the stress-controlled mode, as the coherence in molecular tumbling is disrupted.

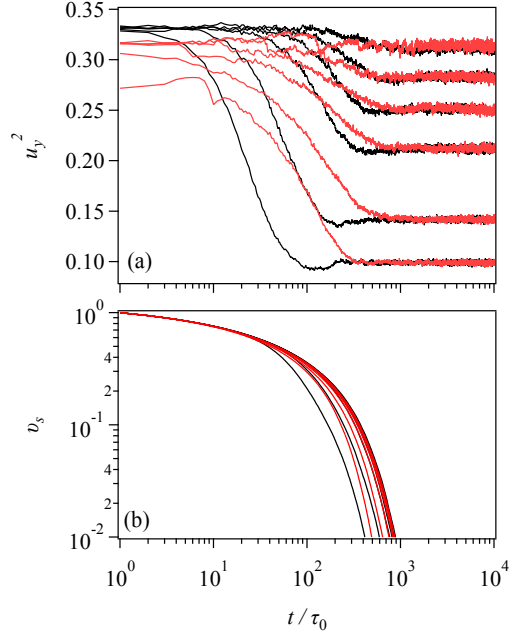

**Figure 7** Time development of square chain end-to-end orientation vector component in the shear-gradient direction $u_y^2$ (a) and number $v_s$ of surviving sliplinks per molecule (b) for stress-controlled (red) and rate-controlled (black) simulations attaining similar steady states. $\dot{\gamma}$ and $\sigma_{xy}$ values at steady state are $(\dot{\gamma}, \sigma_{xy})$ =(0.093, 0.35), (0.037, 0.28), (0.013, 0.21), (0.0077, 0.17), (0.0047, 0.13), and (0.0024, 0.08), from left to right, respectively.

CONCLUSIONS

The primitive chain network simulation was extended to describe creep experiments and tested against literature data on an entangled polybutadiene (PB) solution [14]. Based on the qualitative agreement with the data, the transient molecular motion was analyzed using the decoupling approximation, in which stress is decomposed into separate contributions due to the entanglement density, strand stretch, and strand orientation. The analysis demonstrated that, for all observed molecular characteristics, the time development is relatively mild for stress-controlled cases compared to rate-controlled ones, when the two deformation modes are compared under conditions with similar steady states. The molecular motion was further analyzed by looking at the end-to-end orientation vector and at surviving sliplinks from the start of the flow. These quantities showed that the coherent molecular tumbling is somewhat disturbed under stress-controlled mode, and consequently, deformation-induced disentanglement is suppressed in stress-controlled cases compared to rate-controlled ones.

Although the results represent the first attempt to analyze the creep behavior of entangled polymers via a molecular model, it is fair to note that the simulation cannot quantitatively capture the data by Ge et al [14]. In this respect, further tests on creep data are required for other systems, including branched polymers and polydisperse systems. The effects of so-far neglected molecular mechanisms, such as finite chain extensibility [22], friction change [20], and wall slip [29], are also worth discussing. Supplementary studies in such directions are ongoing, and the results will be reported elsewhere.